\documentclass[a4paper,prl,reprint,showpacs]{revtex4-1}
\usepackage{amsmath}
\usepackage{graphicx}
\usepackage[english]{babel}
\usepackage{subfig}

\begin{document}

\newcommand{\non}{\nonumber}
\newcommand{\be}{\begin{equation}}
\newcommand{\ee}{\end{equation}}
\newcommand{\ba}{\begin{eqnarray}}
\newcommand{\ea}{\end{eqnarray}}
\newcommand{\p}{\partial}
\newcommand{\lb}{\left (}
\newcommand{\rb}{\right )}
\newcommand{\lsq}{\left [}
\newcommand{\rsq}{\right ]}
\newcommand{\lc}{\left \{}
\newcommand{\rc}{\right \}}
\newcommand{\phint}{\int_{0}^{2 \pi} \hspace{-0.2cm}}
\newcommand{\rint}{\int_{0}^{\infty} \hspace{-0.2cm}}
\newcommand{\intv}{\int \hspace{-0.1cm}}
\newcommand{\sinc}{\mathrm{sinc}}

\title{Full characterisation of the quantum spiral bandwidth of entangled biphotons}
\author{Filippo M. Miatto}
\author{Alison M. Yao}
\author{Stephen M. Barnett}
\affiliation{SUPA and Department of Physics, University of
Strathclyde, Glasgow G4 0NG, Scotland, U.K.}

\date{\today}

\begin{abstract}
Spontaneous parametric down conversion has been shown to be a reliable source of entangled photons. Amongst the wide range of properties that have been shown to be entangled, it is the orbital angular momentum that is the focus of our study. We investigate, in particular, the bi-photon state generated using a Gaussian pump beam. We derive an expression for the simultaneous correlations in the orbital angular momentum ($\ell$) and radial momentum ($p$) of the down-converted Laguerre-Gaussian beams. Our result allows us, for example, to calculate the spiral bandwidth with no restriction on the geometry of the beams: $\ell$, $p$ and the beam widths are all free parameters. Moreover, we show that, with the usual paraxial and collinear approximations, a fully analytic expression for the correlations can be derived.
\end{abstract}


\maketitle

\section{Introduction}

Entanglement is one of the most remarkable phenomena introduced by quantum physics. It has applications in quantum 
imaging \cite{DAngelo04,Bennink04,Pittman95,Kolobov07} and quantum cryptographic systems \cite{Ursin07} and is of great interest 
both within the context of quantum information theory \cite{Plenio07} and in quantum computing \cite{Nielsen2000,Barnett2009}.
More recently, entanglement has been demonstrated  between spatial modes carrying orbital angular momentum (OAM) and 
this has been used for quantum information protocols \cite{Mair01,Vaziri02, Groblacher06,Yao06} and to demonstrate violation 
of Bell's inequality \cite{Aiello05,Jack10} and in an angular demonstration of the EPR paradox \cite{Leach10}. Entanglement has also
been shown in the spatiotemporal structure of the light, both for type-I \cite{Gatti09} and for type-II \cite{Brambilla10} parametric down-conversion.

Spatial modes can also carry radial momentum. Both angular ($\ell$) and radial ($p$) momenta can be controlled and converted between one another using diffractive optical elements tipically in the form of spatial light modulators (SLMs) displaying holograms \cite{Leach05,Arlt10}. Recent applications of controlling the entanglement in the radial modes involve topological knots of phase singularities. These can be produced by engineering specific superpositions of radial and orbital angular momentum modes \cite{Leach05}. Topological knots have been shown to be robust against perturbations \cite{Jaqui10}, a feature that might be helpful when propagating entangled states through the atmosphere. Another application of the entanglement in the radial modes is a larger number of qubits that could be used in a quantum information task.

In spontaneous parametric down-conversion (SPDC), a pair of lower-frequency photons are created when a pump field interacts with a nonlinear 
crystal \cite{Hong85,Mair01}. The spatial structure of the down-converted photon states can be expressed as a mode decomposition of their joint wave function in an appropriate basis.
The degree of entanglement is directly related to the number of modes participating in the state, often referred to as the spiral bandwidth \cite{Torres03,Oemrawsingh05}. The strength of the entanglement between pairs of photons can be verified in the laboratory by converting with SLMs specific $\ell$ modes to the fundamental mode of the single mode detection fibers and measuring the coincidence rate \cite{Jaqui10}.

For applications in quantum information, such as multidimensional quantum imaging \cite{Pittman96,Kolobov07}, or for applications involving SLMs, which affect both the $\ell$ and the $p$ quantum numbers \cite{Arlt10,Ren04}, it is important to be able to calculate the exact form of the down-converted biphotons, with no restrictions on any of the defining parameters: the OAM indices $\ell_i,\,\ell_s$, the radial indices $p_i,\,p_s$ or the beam widths $w_p,\ w_i,\ w_s$.

In this paper we consider a typical SPDC set-up as shown in Fig.\ref{fig:setup}. We first analyze the general problem, making use of the usual assumptions for the phase matching (i.e. we assume that the transverse phase matching is satisfied and that there is no extra longitudinal phase mismatch, other than the difference of the $\mathbf k$ vectors of the down-converted photons) and derive an integral solution for the mode amplitudes. This is simple enough to evaluate numerically and we show that it reduces to an analytical solution in the (experimentally relevant) limit of a short crystal. Special cases of our results are in excellent agreement with previous calculations \cite{Torres03}, when the relevant restrictions are applied. 
We show, moreover, that the same analytical result is obtained, without the need for a thin crystal approximation, when we consider collinear geometry.

We use our results to investigate the influence of the relative beam widths and the effect of the radial contributions on the resulting spiral bandwidth.  
We show that the various mode amplitudes have a strong dependence on the ratios of the signal and idler widths to the pump width.
In a future paper we will discuss the effect of the crystal length, phase matching, and the differences between collinear and non-collinear geometries.

\section{General amplitudes of the SPDC state}
\subsection{Geometry and notation}
\begin{figure}[h!]
\includegraphics[width=.45\textwidth]{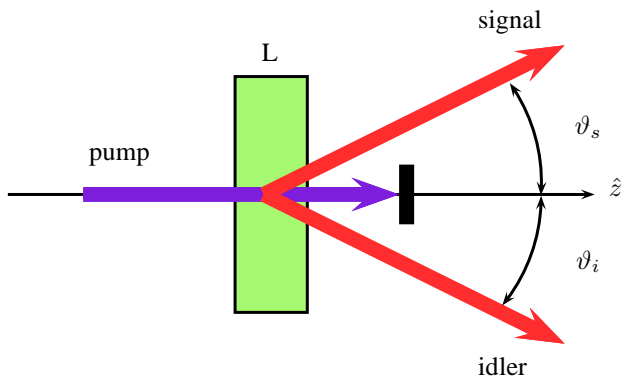}
\caption{(Color online) Sketch of the SPCD process. A gaussian pump propagating in the $z$ direction is incident on a short nonlinear 
crystal. Signal and idler photons are produced at angles $\vartheta_s$ and $\vartheta_i$ to the pump direction.}
\label{fig:setup}
\end{figure}

A typical SPDC setup consists of a continuous-wave Gaussian pump beam propagating in the $z$ direction, as shown in Fig.\ref{fig:setup}, incident on a short (typically $1$--$3$mm) nonlinear crystal of length $L$. This produces two highly-correlated, 
lower-frequency photons, commonly termed signal and idler.  Energy 
is conserved in this process so that $\omega_p = \omega_s + \omega_i$, where the subscripts $p, s, i$ refer to the pump, signal and idler, 
respectively. The photons are emitted at angles $\vartheta_{s,i}$ to the direction of propagation of the pump, $\mathbf{\hat z}$, and the components $\mathbf q_{s,i}$ (perpendicular to the $z$ axis) of their $\mathbf k_{s,i}$ vectors are at angles $\varphi_s$ and $\varphi_i$ to the $x$ axis. This means that $\varphi$ is an azimuthal coordinate and this will play a central role in the description of the orbital angular momentum. The $\mathbf q$ components can be decomposed, on the planes perpendicular to the $z$ axis, as 
\be
\mathbf{q_{s,i}}=\left(\begin{array}{c}
\rho_{s,i} \cos \varphi_{s,i} \\
\rho_{s,i} \sin\varphi_{s,i}\\
0\end{array}\right) 
\ee
where the radial variable, $\rho$, extends outwards from $\mathbf{\hat z}$.
The magnitude of the wave-vector inside the medium is $\omega n/c$ and for type-I phase matching (eoo) the extra-ordinary and ordinary refractive indices are $n_e(\omega_p)$ and $n_o(\omega_{s,i})$. Typical values, for example for $\beta$-barium borate (BBO), are: at $\lambda=1064$nm, $n_o = 1.65, n_e = 1.54$ and at $\lambda=532$nm, $n_o = 1.67, n_e = 1.55$ \cite{Nikogosyan}.

At the output of the nonlinear crystal the two-photon state in the wave-vector domain is given by \cite{Saleh00,Torres03}
\be
| \psi \rangle = \intv \intv  d\mathbf{k}_s d\mathbf{k}_i \Phi( \mathbf{k}_s, \mathbf{k}_i) \hat{a}_s^{\dagger}( \mathbf{k}_s) \hat{a}_i^{\dagger}( \mathbf{k}_i) | 0 \rangle ,
\ee
where $\Phi( \mathbf{k}_s,  \mathbf{k}_i)$ describes the mode function of the pump and the phase matching conditions, $| 0 \rangle$ is the multimode vacuum state and $\hat{a}_s^{\dagger}( \mathbf{k}_s), \hat{a}_i^{\dagger}( \mathbf{k}_i)$ are creation operators for the signal and idler modes with wave vectors $ \mathbf{k}_s,  \mathbf{k}_i$, respectively.

Photon pairs generated by parametric down-conversion are entangled in arbitrary superpositions of OAM modes and we aim to complete this description by including the radial behaviour.
A natural way to do this is to describe the down-converted photons in terms of the Laguerre-Gaussian (LG) modes, $LG_p^{\ell}$. Here $\ell$ corresponds to the angular momentum carried by the mode, $\ell \hbar$, and describes the helical structure of the wave front around a wave front singularity and $p$ is the number of radial zero crossings.\\

\subsection{Calculation of the coincidence amplitudes}
The coincidence probability for finding one signal photon in a given LG mode characterized by the $\ell_s$ and $p_s$ numbers and one idler photon in a given LG
mode characterized by the $\ell_i$ and $p_i$ numbers is $P_{p_s,p_i}^{\ell_s,\ell_i} = \left |C_{p_s,p_i}^{\ell_s,\ell_i} \right|^2$, where the coincidence amplitudes
$C^{\ell_s,\ell_i}_{p_s,p_i}$ are given by the overlap integral 
\ba
&&C_{p_s,p_i}^{\ell_s,\ell_i} = \langle \psi_i,\psi_s|\psi_{SPDC} \rangle \nonumber \\
&&= \intv \intv d^{3}k_s d^{3}k_i \, \Phi(\mathbf{k}_s, \mathbf{k}_i) 
\lsq LG_{p_s}^{\ell_s}(\mathbf{k}_s) \rsq^* 
\lsq LG_{p_i}^{\ell_i}(\mathbf{k}_i) \rsq^*
\label{eqn:overlap1}
\ea
The pump profile and phase matching condition are given by \cite{Saleh00}
\be
\label{pumpphasematching}
\Phi(\mathbf{k}_s, \mathbf{k}_i) = \intv d^{3}k_p \tilde{E}_p(\mathbf{k}_p) \xi({\mathbf{k}_p - \mathbf{k}_s - \mathbf{k}_i}) \delta(\omega - \omega_s - \omega_i)
\ee
where the $\delta$ term enforces energy conservation and the $\xi$ term arises due to phase matching.
We can simplify this if, instead of the wave vector $\mathbf{k}$, we use its transverse component $\mathbf q$ and the 
corresponding frequency $\omega$. As we have assumed a monochromatic gaussian pump of frequency 
$\omega_p=\omega_i+\omega_s$ we can write
\be
\tilde{E}_p(\mathbf{k}_p) = \tilde{E}_p(q) \delta(\omega - \omega_p)
\ee
where $\tilde{E}_p(q)$ is the Fourier transform of 
\be
E_p(r,\phi) = \sqrt{\frac{2}{\pi}} \frac{1}{w_p} \exp \left( \frac{- r^2}{w_p^2} \right) \, ,
\ee
the spatial distribution of the pump at the input face of the crystal.

For a crystal of finite thickness, $L$, in the longitudinal direction and transverse length much larger than the pump beam size, the phase matching condition is \cite{Saleh00}
\ba
\xi({\mathbf{k}_p - \mathbf{k}_s - \mathbf{k}_i}) &=& \delta(q_p - q_s - q_i) \nonumber \\
&\times& \sqrt{\frac{L}{\pi k_p}} \, \sinc \lb \frac{L \Delta k_{z}}{2} \rb \exp \lb\frac{-i L \Delta k_{z}}{2}  \rb \nonumber 
\ea
where $\Delta k_{z} = k_{p,z} - k_{s,z} - k_{i,z}$ and $k_{z}= \sqrt{n^2(\omega) \omega^2/c^2 - q^2}$ is the longitudinal 
component of the wave vector, $\mathbf{k}$, with transverse component $q$, angular frequency $\omega$ 
and refractive index $n(\omega)$.

If the angle between signal and idler beams is small enough that the $z$-component of the momentum vector ($\sqrt{k^2-q^2})$ can be approximated by $k-q^2/2k$, where $q=|\mathbf{q}|$, we can write the phase matching function as
\ba
\Phi(\mathbf{k}_s, \mathbf{k}_i) = \tilde{E}(q_s + q_i) \sqrt{\frac{2L}{\pi^{2}k_{p}}} \, \sinc \lb \frac{L \Delta k}{2} \rb e^{-i\frac{L \Delta k}{2}} 
\ea
where 
\ba
\Delta k &=&\frac{|\mathbf{q}_i-\mathbf{q}_s|^2}{2k_p}.
\ea
Explicitly:
\begin{align}
\Phi(\mathbf{q}_s,\mathbf{q}_i)&=\overbrace{\frac{w_{p}}{\sqrt{2\pi}}e^{-\frac{w_{p}^{2}}{4}|\mathbf{q}_s+\mathbf{q}_i|^{2}}}^{\mathrm{Pump}}\times\nonumber\\
&\times\underbrace{\sqrt{\frac{2L}{\pi^{2}k_{p}}}\sinc \lb \frac{L |\mathbf{q}_i-\mathbf{q}_s|^2}{4k_p} \rb e^{-i\frac{L |\mathbf{q}_i-\mathbf{q}_s|^2}{4k_p}}}_{\mathrm{Phase\ Matching}}
\label{pump+phm}
\end{align}

It is convenient to calculate the coincidence amplitudes in a cylindrical coordinate system, so we re-express (\ref{eqn:overlap1}) as:
\ba
 C_{p_s,p_i}^{\ell_s,\ell_i} &\propto& \rint \rint \phint \phint \Phi(\rho_{i},\rho_{s},\varphi_{i},\varphi_{s}) 
 \left [ LG_{p_s}^{\ell_s}(\rho_{s},\varphi_{s})\right]^{*} \nonumber \\
&\times& 
\left [ LG_{p_{i}}^{\ell_i}(\rho_{i},\varphi_{i})\right]^{*}
\rho_{i}\rho_{s}\,d\rho_{i}d\rho_{s}d\varphi_{i}d\varphi_{s},
\label{overlap}
\ea
where $\rho$ and $\varphi$ are the modulus and azimuthal angle, respectively, of the transverse component $\mathbf q$ of the wave vector and 
the normalized LG modes in $k$-space are given by
\ba
LG_p^{\ell}(\rho,\varphi)&=& \sqrt{\frac{w^2 p!}{2 \pi  \left( p+| \ell | \right)!}} \left( \frac{\rho w}{\sqrt{2}} \right)^{| \ell |} 
\exp \lb \frac{- \rho^2 w^2}{4} \rb  \nonumber \\
\times &(-1)^p&   L_p^{| \ell |}  \left( \frac{\rho^2 w^2}{2} \right) \exp \left [i \ell \lb \varphi + \frac{\pi}{2} \rb \right ] .
\ea
Here $w$ is the beam waist (we have assumed $z=0$) and $L_p^{| \ell |}(\cdot)$ is an associated Laguerre polynomial.

The pump and phase matching functions in \eqref{pump+phm} can be written in cylindrical coordinates by performing the substitution 
\begin{align}
|\mathbf q_{i}\pm \mathbf q_{s}|^{2}=\rho_{i}^{2}+\rho_{s}^{2}\pm2\rho_{i}\rho_{s}\cos(\varphi_{i}-\varphi_{s})
\end{align}
to obtain
\begin{widetext}
\begin{align}
\Phi(\rho_i,\rho_s,\varphi_i,\varphi_s)=\frac{w_{p}}{\sqrt{2\pi}}e^{-\frac{w_{p}^{2}}{4}(\rho_{i}^{2}+\rho_{s}^{2}+2\rho_{i}\rho_{s}\cos(\varphi_{i}-\varphi_{s}))}
\sqrt{\frac{2L}{\pi^{2}k_{p}}}\sinc \lb L\frac{\rho_{i}^{2}+\rho_{s}^{2}-2\rho_{i}\rho_{s}\cos(\varphi_{i}-\varphi_{s})}{4k_p} \rb e^{-iL\frac{ \rho_{i}^{2}+\rho_{s}^{2}-2\rho_{i}\rho_{s}\cos(\varphi_{i}-\varphi_{s})}{4k_p}}
\end{align}
\end{widetext}

It is then straightforward to see that $\Phi(\rho_i,\rho_s,\varphi_i,\varphi_s)$ depends on the radial coordinates in the momentum space, $\rho_i$ and $\rho_s$, and on the \emph{difference} between the azimuthal angles, $\varphi_i-\varphi_s$.
This allows the function to be written as a superposition of plane waves with phase $\exp(i\ell(\varphi_i-\varphi_s))$:
\begin{align}
\Phi(\rho_i,\rho_s,\varphi_i-\varphi_s)=\sum_{\ell=-\infty}^{\infty} f_\ell(\rho_i,\rho_s)e^{i\ell(\varphi_i-\varphi_s)} .
\end{align}
Using this in (\ref{overlap}),  
the angular integral becomes
\begin{eqnarray}
\sum_{\ell=-\infty}^{\infty} f_\ell(\rho_i,&&\rho_s)\int_0^{2\pi}\int_0^{2\pi} e^{i\ell_s\varphi_s} e^{i\ell_i\varphi_i}  e^{i\ell(\varphi_i-\varphi_s)}d\varphi_i\,d\varphi_s \nonumber\\
&& \propto \delta_{\ell,-\ell_i}\delta_{\ell,\ell_s},
\end{eqnarray}
which clearly enforces the angular momentum conservation, $\ell_i+\ell_s=0$.

We also re-write the $\sinc$ function
as the inverse Fourier transform of the step function:
\be
\sinc \lb \frac{|\mathbf{q}_i - \mathbf{q}_s|^2 L}{4 k_p} \rb = \frac{1}{L} \int_{-L/2}^{L/2} \hspace{-0.3cm} dt \exp \lb -\frac{i |\mathbf{q}_i - \mathbf{q}_s|^2 t}{2 k_p} \rb .
\ee
In this way we calculate
\ba
C_{p_{i},p_{s}}^{{\ell,-\ell}} && \propto K_{p_i,p_s}^{|\ell|}\int_{-L/2}^{L/2}  \hspace{-0.3cm} dt \, \frac{B^{|\ell|}(1-\frac{4I}{T})^{p_{s}}(1-\frac{4S}{T})^{p_{i}}}{T^{|\ell|+1}} \nonumber \\
\times && \ _{2}F_{1}\left[{-p_{i},-p_{s}\atop-p_{i}-p_{s}-|\ell|};\frac{T(T-4I-4S+4)}{(T-4S)(A-4I)}\right]
\label{eqn:NumSpiral}
\ea
where the combinatorial coefficient $K_{p_i,p_s}^{|\ell|}$ is given by:
\ba
K_{p_i,p_s}^{|\ell|}=\frac{(p_i+p_s+|\ell|)!}{\sqrt{p_i!p_s!(p_s+|\ell|)!(p_i+|\ell|)!}} .
\ea
$B$, $I$, $S$ and $T$ are functions of the dummy variable $t$:
\ba
B \, &&=-\left(\frac{2t}{w_{i}w_{s}k_{p}}+\frac{L}{w_{i}w_{s}k_{p}}+\frac{iw_{p}^{2}}{w_{i}w_{s}}\right)  \nonumber \\
T \, &&=4IS+B^{2} \\
I \, &&=\frac{w_{p}^{2}}{2w_{i}^{2}}+\frac{1}{2}+\frac{it}{w_{i}^{2}k_{p}}+\frac{iL}{2w_{i}^{2}k_{p}}  \nonumber \\
S \, &&=\frac{w_{p}^{2}}{2w_{s}^{2}}+\frac{1}{2}+\frac{it}{w_{s}^{2}k_{p}}+\frac{iL}{2w_{s}^{2}k_{p}}  \nonumber
\ea
and $_2F_1$ is the Gauss hypergeometric function. Although this integral is too complicated to be calculated analytically, it is simple enough to be evaluated numerically.

In the limit of a thin crystal, however, we can solve the integral analytically (because the integration limits depend on the crystal length). 
This gives
\begin{align}
C_{p_{i},p_{s}}^{{\ell,-\ell}} & \propto K_{p_{i},p_{s}}^{|\ell|}\frac{(1-\gamma_i^2+\gamma_s^2)^{p_{i}}(1+\gamma_i^2-\gamma_s^2)^{p_{s}}(-2\gamma_i\gamma_s)^{|\ell|}}{(1+\gamma_i^2+\gamma_s^2)^{p_{i}+p_{s}+|\ell|}}\nonumber \\
& \times {}_{2}F_{1}\left[{-p_{i},-p_{s}\atop-p_{i}-p_{s}-|\ell|};\frac{1-(\gamma_i^2+\gamma_s^2)^{2}}{1-(\gamma_i^2-\gamma_s^2)^{2}}\right]
\label{eqn:ShortSpiral}
\end{align}
where $\gamma_i$ and $\gamma_s$ are the ratios $w_p/w_i$ and $w_p/w_s$, so that the two $\gamma$ factors are the inverse signal and idler widths normalized to the pump width. This means that every amplitude is invariant under the joint scaling of signal, idler and pump widths.

Comparing (\ref{eqn:ShortSpiral}) with the numerical evaluation of (\ref{eqn:NumSpiral}) we find excellent agreement for crystal lengths up to tens of centimeters, which
is two orders of magnitude larger than the crystals typically used in experiments. Having a thick crystal, therefore, will not greatly affect the mode content of the SPDC state, as long as the conditions of small emission angles, and similar refractive indices for pump, signal and idler fields, which allow $\Delta k_z\simeq|\mathbf q_i-\mathbf q_s|^2/2k_p$, are met.

\subsection{Relation between collinear geometry and thin crystals}
Note that in collinear geometries ($\mathbf{q}_s = \mathbf{q}_i=0$) the $\sinc$ term in the phase matching condition is unity. This creates a similarity between having a thin crystal or tuning a thicker crystal to reach collinear conditions. The biphoton state in this case is given by \cite{Saleh00,Torres03b}
\be
|\psi\rangle = \int dr \Phi(r) \hat{a}_s^{\dagger}(r)  \hat{a}_i^{\dagger}(r) |0\rangle
\ee
where $\Phi(r)$ is the spatial distribution of the pump beam at the input face of the crystal, and $r$ is the radial coordinate in real space.
The coincidence amplitudes can then be calculated readily from 
\ba
C_{p_s,p_i}^{\ell_s,\ell_i} &\propto& \phint d\phi \rint r \, dr  \nonumber \\
&\times& LG_{0}^{0}(r, \phi) \lsq LG_{p_s}^{l_s}(r, \phi) \rsq^* \lsq LG_{p_i}^{l_i}(r, \phi) \rsq^*
\label{eqn:ColSpiral}
\ea
where we use the Laguerre-Gaussian modes in the real-space form:\\
\ba
LG_p^{\ell} (r, \phi)&=&\sqrt{\frac{2 p!}{\pi \left( p+| \ell | \right)!}} \frac{1}{w} \left( \frac{r \sqrt{2}}{w} \right)^{| \ell |} 
\exp \left( \frac{- r^2}{w^2} \right)  \nonumber \\
&\times& L_p^{| \ell |}\left( \frac{2 r^2}{w^2} \right) \exp(i \ell \phi) .
\ea
In this case the integral over the azimuthal coordinate is
\be
\phint \, \, d \phi \exp \lsq -i (\ell_s + \ell_i) \phi \rsq = 2 \pi \delta_{\ell_s,-\ell_i}
\ee
from which we obtain the well-known conservation law for the orbital angular momentum: $\ell_s = -\ell_i$ \cite{FrankeArnold02}.

Substituting these into (\ref{eqn:ColSpiral}) we find exactly the same expression as (\ref{eqn:ShortSpiral}), which means that the amplitudes for a collinear beam and a crystal of any experimentally realisable size are the same for a quasi-collinear beam in the thin crystal approximation.\\

\section{Structure of the SPDC state}
The advantage of an analytical result over an integral one, such as \eqref{eqn:NumSpiral}, is that it allows us to see more easily the role that each parameter plays in determining 
the state of the down-converted photon and the resulting $\ell$ distribution, or spiral bandwidth \cite{Torres03}.
Full knowledge of this allows the quality of the entangled state to be determined and also allows us to compare the width of the 
distribution of $\ell$ modes of different states.
From (\ref{eqn:ShortSpiral}) it is clear that the distribution of modes is determined by the ratio of the signal and idler widths to the pump width. We can see this more clearly, and also investigate more the effect of taking the radial modes into account, if we consider some special cases.\\
\subsection{Comparison with published results}
\begin{figure}[h!]
\includegraphics[width=0.48\textwidth]{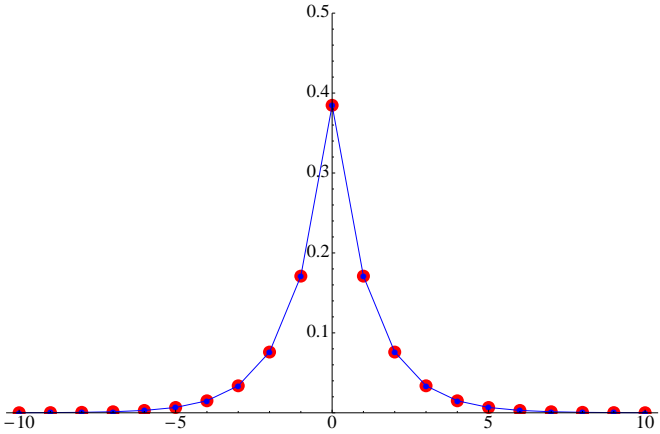}
\caption{(Color online) Normalized spiral bandwidth for $\gamma_s = \gamma_i = 1$ and 
$p_s=p_i=0$. Results from eq. (10) in \cite{Torres03} are in red.}
\label{fig:CompTor}
\end{figure}

Note that for the specific case of $p_s = p_i = 0$ eq.(\ref{eqn:ShortSpiral}) reduces to
\ba
|C_{0,0}^{\ell,-\ell}|^2 \propto  \lb \frac{2\gamma_i\gamma_s}{1+\gamma_i^2+\gamma_s^2} \rb^{ 2| \ell |} .
\ea
If we further choose the signal and idler widths to be equal, so that $\gamma_i=\gamma_s=\gamma$, then this simplifies to
\ba
|C_{0,0}^{\ell,-\ell}|^2 \propto\lb \frac{2\gamma^2}{1+2\gamma^2} \rb^{2| \ell |} .
\label{eqn:PartCase}
\ea
This agrees with the result found in \cite{Torres03} as depicted in Fig.\ref{fig:CompTor}:\\

\ \\

Although the form of eq.\eqref{eqn:PartCase} is very simple, it is remarkably precise when it is compared to the numerical evaluation of eq.(\ref{eqn:NumSpiral}), where the signal and idler sizes are equal and if the eigenstates of $p_i=p_s=0$ have been selected.
\subsection{Correlation between $p_i$ and $p_s$}

It is possible to have an extended view of the structure of the entangled system with an array plot that shows the contribution $P^{\ell,-\ell}_{p_s,p_i}=|C^{\ell,-\ell}_{p_s,p_i}|^2$ for each pair of modes.\\
\subsubsection{(i) Effect of the pump width on the correlation between $p_i$ and $p_s$ when $\ell=0$ and signal and idler have the same size.}
\begin{figure}[ht]
\centering
\includegraphics[width=.45\textwidth]{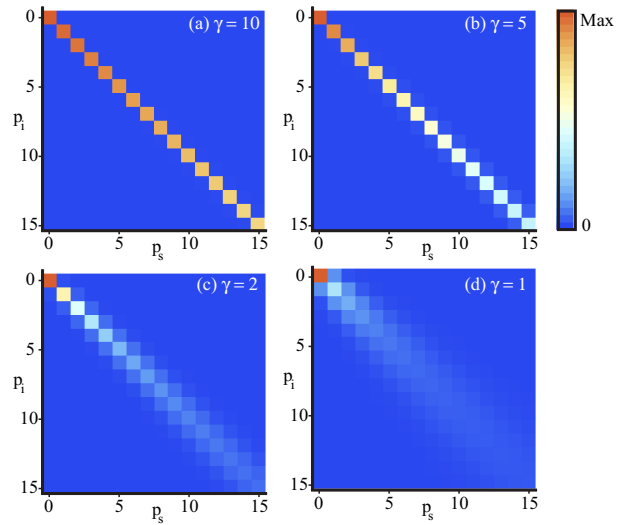}
\caption{(Color online) $P^{0,0}_{p_s,p_i}$ for $0<p_i<15$ and $0<p_s<15$, for different pump sizes. The color bar shows the color normalisation, which takes place between the minimum (0) and the maximum values in each graph. The same happens in Fig.\ref{fig:ppcorr2} and Fig.\ref{fig:LG}.}
\label{fig:ppcorr}
\end{figure}

We show in Fig.\ref{fig:ppcorr} the contributions of modes in the form $|p_i,0\rangle\otimes|p_s,0\rangle$. Since $\gamma_i=\gamma_s$ in this case, we will omit the subscript and simply write $\gamma$. On the axes we are scanning the discrete values of $p_i$ and $p_s$.

In the limit of an infinite pump width $p_i$ and $p_s$ are delta correlated. Notice how the correlation between $p_i$ and $p_s$ breaks down as the pump beam size approaches that of signal and idler. (This corresponds to $\gamma \rightarrow 1$).\\

In an experiment it is not possible to use a pump with arbitrarily large beam waist because the crystal size places an effective upper limit. It is also not possible to shrink the size of signal and idler enough so that $\gamma >>1$. Therefore, the experimental scenario will be closer to the bottom right graph rather than the top left. This means that cross correlations between eigenstates of different $p$ are to be expected.

\subsubsection{(ii) Effect of $\ell$ on the correlation between $p_i$ and $p_s$.}

In the previous section we described the $p$ correlations between states of OAM with $\ell=0$. We now consider the case $\ell\neq0$.

It is worth recalling that (in general) the probability of selecting a joint state of a given OAM, $|\ell|$, decreases as the value of $|\ell|$ increases. 
For this reason, if we select a specific value of $\ell$, as is done with SLMs, the average rate of coincidence counts will be lower than the rate measured after selecting states with a lower value of $|\ell|$. If the value of the selected $|\ell|$ is too large, the rate could drop too much to obtain a meaningful signal to noise ratio and the effects could be masked by noise. This is why we show the effect only up to the value $|\ell|=6$.

\begin{figure}[h!]
\centering
\includegraphics[width=.45\textwidth]{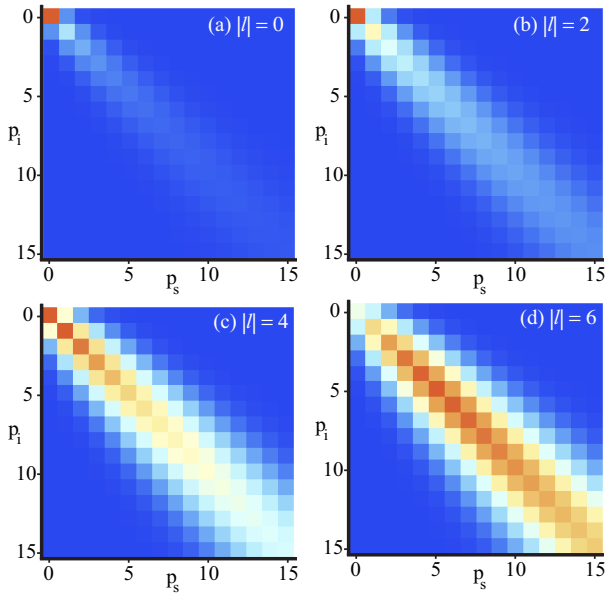}
\caption{(Color online) Correlations between $0<p_i<15$ and $0<p_s<15$, for different $|\ell|$ values and a pump of the same size as  signal and idler. Notice how the combination of $p_i$ and $p_s$ with the highest probability of being measured shifts along the diagonal as higher and higher values of $|\ell|$ are selected. Each graph is normalised to the maximum value, a comparison between the main diagonals for $|\ell|=4,5,6,7$ is given in Fig.\ref{fig:overlap}.}
\label{fig:ppcorr2}
\end{figure}
A physical reason why the maximum probability of detecting modes of similar $p$ shifts to a higher value of $p$ for states of a higher OAM eigenvalue can be found by considering what happens to the product of the pump mode with specific signal and idler modes.
In this case the key idea is that when we calculate the overlap integral between the pump and modes with a given $\ell$, there is an optimal combination of $p_s$ and $p_i$ that maximizes the overlap. The higher the value of $\ell$, the higher the optimal value of the combination of radial indices.

\begin{figure}[h!]
\centering
\includegraphics[width=.47\textwidth]{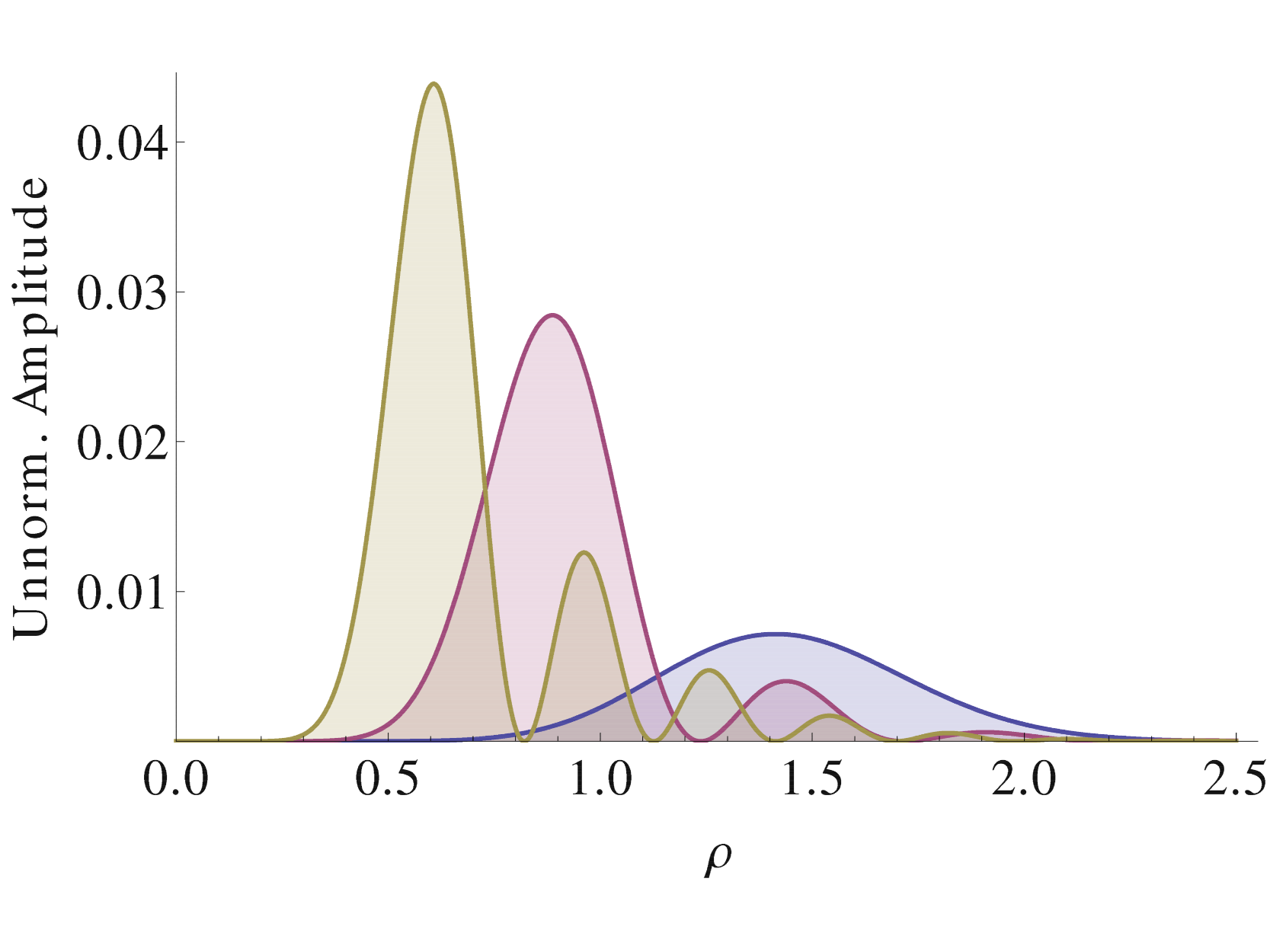}
\caption{(Color online) Three products in the form $LG_0^0LG_p^\ell LG_p^{-\ell}$ with $\ell=6$ and $p=0,5,15$ (respectively in blue, red and brown), plotted radially from the origin of the cylindrical coordinate system.}
\label{fig:LGG}
\end{figure}

\begin{figure}[h!]
\centering
\includegraphics[width=.47\textwidth]{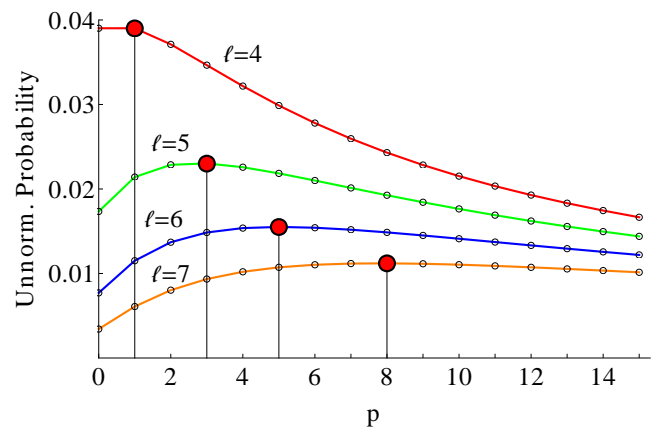}
\caption{(Color online) The overlap between $LG_0^0$ and $LG_p^{\ell}LG_p^{-\ell}$ is best achieved at an optimal value of $p$ (indicated by red disks), once $|\ell|$ is fixed. The red line represents the main diagonal in the plot in Fig.4c and the blue line represents the main diagonal in the plot in Fig.4d.}
\label{fig:overlap}
\end{figure}

To show how this happens, consider the product of the fundamental pump mode with two $p_i=p_s$ modes (i.e. the integrand of \eqref{overlap} in the short crystal limit). In Fig.\ref{fig:LGG} we show three such products for $p_i=p_s=0,5,15$, as functions of $\rho$, omitting the angular dependence.
If we calculate the (unnormalised) overlap integral, it is possible to see (Fig.\ref{fig:LGG}) that, although the effect of increasing the radial indices moves the inner rings towards the origin of the coordinates (where the maximum value of the gaussian mode of the pump is), the radius of the main ring reduces and consequently the overlap will reduce. Thus, as the number of rings increases, an optimal value of the overlap integral is reached and subsequently it drops, which is exactly what happens along the diagonals of the graphs in Fig.\ref{fig:ppcorr2}, which are also shown in Fig.\ref{fig:overlap}.
\subsubsection{(iii) Effect of signal-idler size mismatch on the correlation between $p_i$ and $p_s$.}
In the analysis of the correlations between states of different $p$ index we assumed the signal and idler fields to have the same size. We now generalize to fields of different size.

When the sizes of signal and idler beams begin to differ, the main correlation line between modes of $p_i$ and $p_s$ shifts towards one or the other axis, depending which of the beams is larger, thus yielding the highest correlations between modes of different radial indices rather than between modes of equal radial indices.

\begin{figure}[h!]
\centering
\includegraphics[width=.45\textwidth]{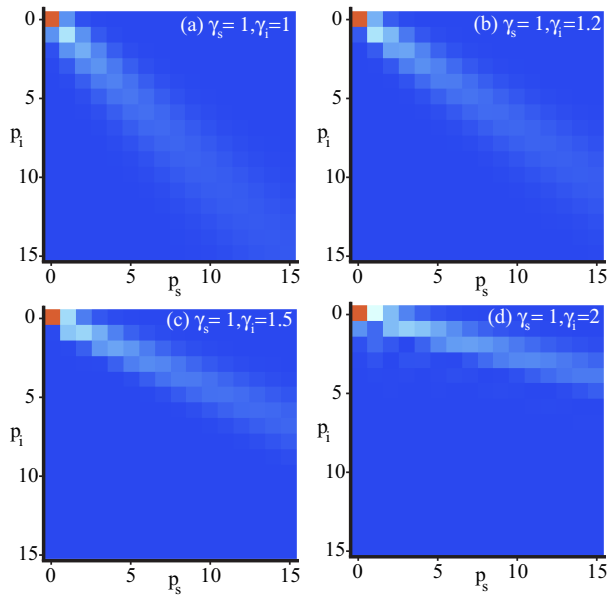}
\caption{(Color online) These graphs show that the effect of a width mismatch yields a higher probability of finding a state with different radial indices. If, instead, the value of $\gamma_s$ were larger than $\gamma_i$ the graphs would be mirrored with respect to the leading diagonal.}
\label{fig:LG}
\end{figure}

Fig.\ref{fig:LG} is a group of four graphs, each featuring the probabilities of detecting an eigenstate in the form $|p_i,0\rangle\otimes|p_s,0\rangle$, where the size of the idler beam increases to twice the size of the signal beam, which remains of the same size of the pump.

It is clear from the graphs that we can increase the detection probability of modes of very different values of $p$ simply by changing the relative size of signal and idler modes.
The consequence of selecting a different $\ell$ mode is analogous to what is shown in Fig.\ref{fig:ppcorr2}.

As we are able to calculate the amplitudes for any values of the beams sizes, it is interesting to see what the effects of the beams sizes are on the spiral bandwidth. We will discuss this in the next section and give multiple examples of the effect of varying the beams sizes.

\subsection{Effect of beams sizes on the spiral bandwidth.}
We now want to see the effects of the beams sizes on the spiral bandwidth (SB). The SB is the collection of detection probabilities in the form $P^{\ell,-\ell}_{p_s,p_i}$, where the radial indices $p_i$ and $p_s$ have been fixed and we scan the $\ell$ eigenvalues in an interval (in our case of $\pm 20$) around $\ell=0$.
Unlike the previous graphs, each graph now \emph{does not} represent the structure of the SPDC state, but rather a collection of ``slices'' of the total modal content, each slice representing the SB, i.e. each graph features only the $\ell$ eigenvalues.

Each group of graphs will highlight a particular feature of the SB determined by the radial indices. Note that the normalization has to be performed on each individual horizontal line and not on the graph as a whole, because the beam size is just a parameter of the detection basis. We stress that each value in the graphs corresponds to the integral over the whole area where the pump overlaps with the detection mode.

We will divide the effects into four families, each one corresponding to equal or different radial indices and equal or different signal-idler sizes. In this way we can cover all the experimentally interesting effects on the SB.
\subsubsection{(i) Equal radial indices - Equal signal-idler sizes.}
This case is of particular importance because now the ratio between pump and signal-idler beams size is the only feature that influences the SPDC state. It is a quantity that is relatively easy to manipulate, for example by choosing an initial pump beam size and then magnifying or de-magnifiying the signal and idler beams. We acknowledge that it can be hard to achieve a particular magnification, because of limiting apertures in the setup, especially between the crystal and the detection instrumentation. Such finite apertures will give a limit on the maximum size of the beams, and also losses if this maximum size is exceeded.

\begin{figure}[h!]
\centering
\includegraphics[width=.45\textwidth]{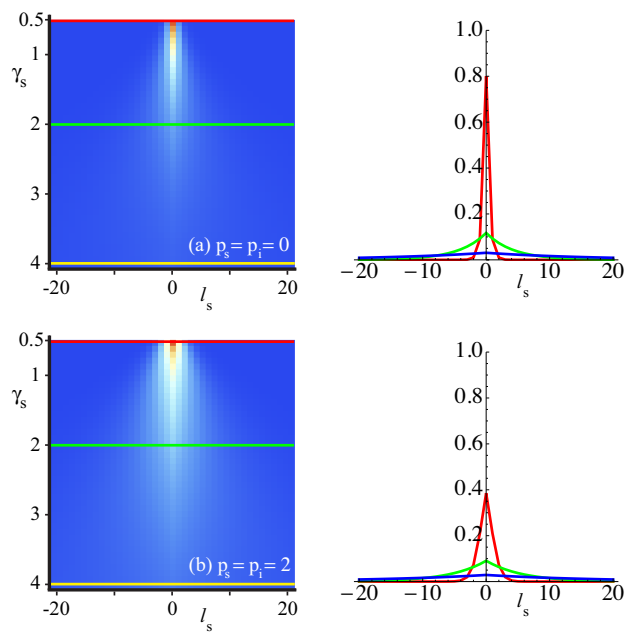}
\caption{(Color online) These graphs show how the SB varies with the pump width, which increases from top to bottom of the graphs and its ratio with the signal and idler beams is shown on the left.
We can see how the radial numbers influence this dependence by comparing the top with the bottom graphs. The positions where the bandwidths that are shown on the right are taken are marked on the graphs on the left (the blue bandwidth on the right is indicated in yellow on the left).}
\label{fig:BAND1}
\end{figure}

The graphs on the left-hand side of Fig.\ref{fig:BAND1} show the SB (horizontally) for different values of the ratio of the pump width to the signal-idler width, i.e. $\gamma$ (which changes on the vertical axis). 

A larger pump width (or smaller signal and idler widths) increases the width of the SB. Also, measuring on a basis with larger radial indices gives a larger SB. In fact, modes with more rings still give a significant overlap with the gaussian pump for larger values of $\ell$ than modes with less rings.

\subsubsection{(ii) Equal radial indices - Different signal-idler sizes.}
\begin{figure}[h!]
\centering
\includegraphics[width=.45\textwidth]{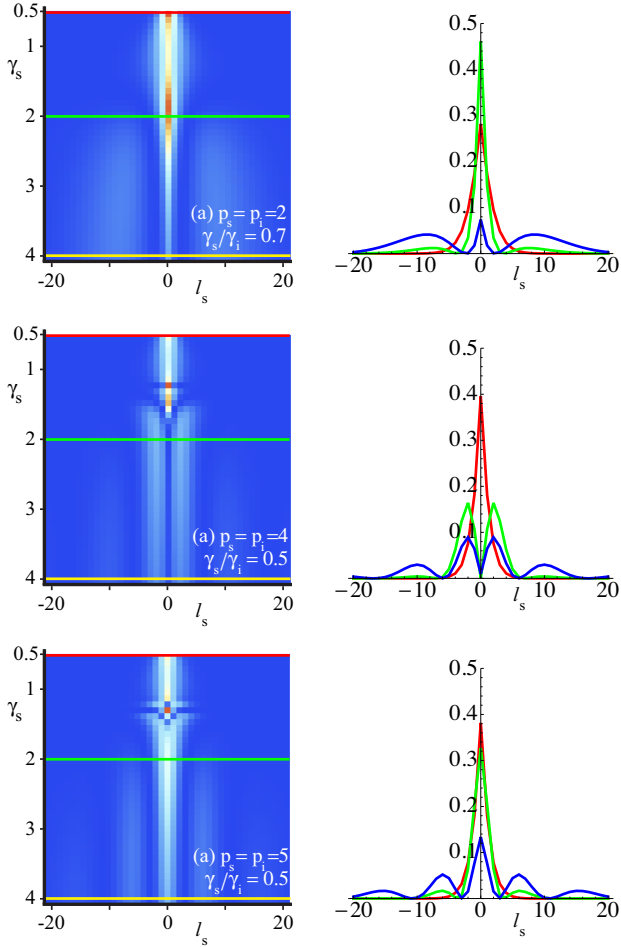}
\caption{(Color online) These graphs show how the difference in the width of signal and idler influences the SB. We show the effect for various values of $p_i=p_s$.}
\label{fig:BAND2}
\end{figure}

A rather different and interesting effect is achieved when the size of signal and idler differ by some amount. Fig.\ref{fig:BAND2} shows similar graphs to Fig.\ref{fig:BAND1}, but here the signal and idler beams now have different widths.

A physical explanation of the multiple-branched SBs in Fig.\ref{fig:BAND2} is given by the fact that the joint detection mode is made out of two modes with different widths (therefore it will consist of concentric rings of positive and negative value) and by the consequence of increasing $\ell$: the rings in the modes move further away from where the pump mode is concentrated, but since they have a different width their product will change shape, becoming alternately mostly negative or mostly positive. The higher the width difference, the higher number of times this process takes place in the same amount of $\ell$ values.

The consequence is that the value of the overlap with the pump mode (which is always positive) increases and decreases alternately as higher and higher $\ell$ modes are chosen, because the mostly positive or mostly negative part of the joint detection mode, since they are moving away from the center of the beam, will cease to have a substantial overlap.

The number of branches in the graphs in Fig.\ref{fig:BAND2} depends on the relative dimensions of signal and idler, and also on the number of rings in the joint detection mode, as mostly positive and negative parts can co-exist in the same detection mode over many rings.  If the pump is too small (top of the graphs on the left-hand side, and red SB in the graphs on the right-hand side) the effect can't be noticed because the gaussian pump always stays mostly inside the first ring (so it doesn't matter now many there are outside the first one, or their values). However, for a larger value of the pump (bottom of the graphs), it initially overlaps with many rings, whose overlap can then be lost, in the way explained, while we consider higher and higher values of $|\ell|$. The blue SB in the right-hand side graphs portrays this effect.

\subsubsection{(iii) Different radial indices - Equal signal-idler sizes.}
We now consider the case of different values of the radial indices $p_i$ and $p_s$. Since the probability of detecting modes of different radial index depends on the ratios of the beam sizes, there are choices of widths that may significantly increase the probability of detecting some particular output modes, as seen in Fig.\ref{fig:LG}.

\begin{figure}[h!]
\centering
\includegraphics[width=.45\textwidth]{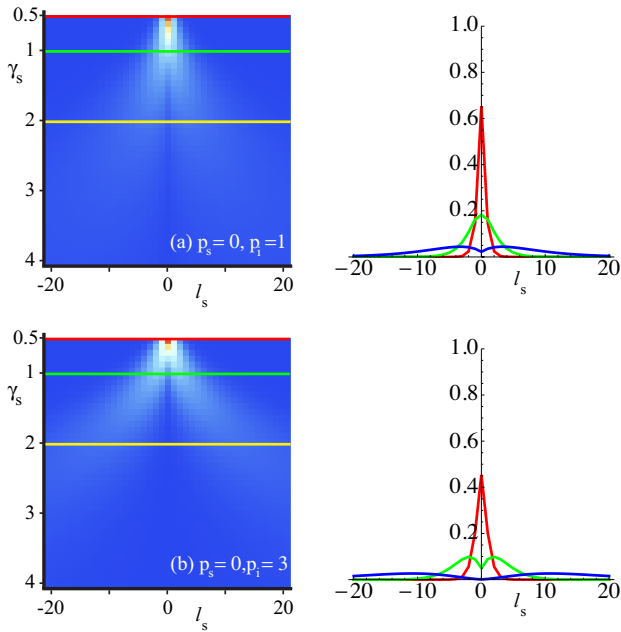}
\caption{(Color online) These graphs show how different radial indices in the signal and idler modes (with equal widths) influence the SB. We show the effect for various values of $p_i$, $p_s$.}
\label{fig:BAND3}
\end{figure}

In Fig.\ref{fig:BAND3} we show the effect of increasing the pump width while maintaining the signal and idler at the same size, but with different radial indices.

As long as signal and idler beams retain the same width, the number of branches will be always two, independent of the values of $p_i$ and $p_s$. This behavior is due to the fact that the beams have the same width, therefore, as higher and higher values of $\ell$ are reached, the rings in the joint detection mode will be moved away from the pump ``rigidly'', i.e. maintaining the same shape and thus delivering a smooth decay of the overlap.

\subsubsection{(iv) Different radial indices - Different signal-idler sizes.}
We showed that a multiple-branched SB can be obtained when signal and idler beams have different sizes, now we show the same effect while also varying $p_i$ and $p_s$. To give a more complete description of the possible SBs we will now consider the \emph{difference} between $\gamma_i$ and $\gamma_s$, rather than the ratio. In the first four graphs the ratios of the pump width with signal and idler widths have a difference (i.e. $\gamma_i-\gamma_s$) of 0.5, 0.75, 1 and 2. The pump width indicated on the left axis of the graphs is relative to the signal width.

\begin{figure}[h!]
\centering
\includegraphics[width=.45\textwidth]{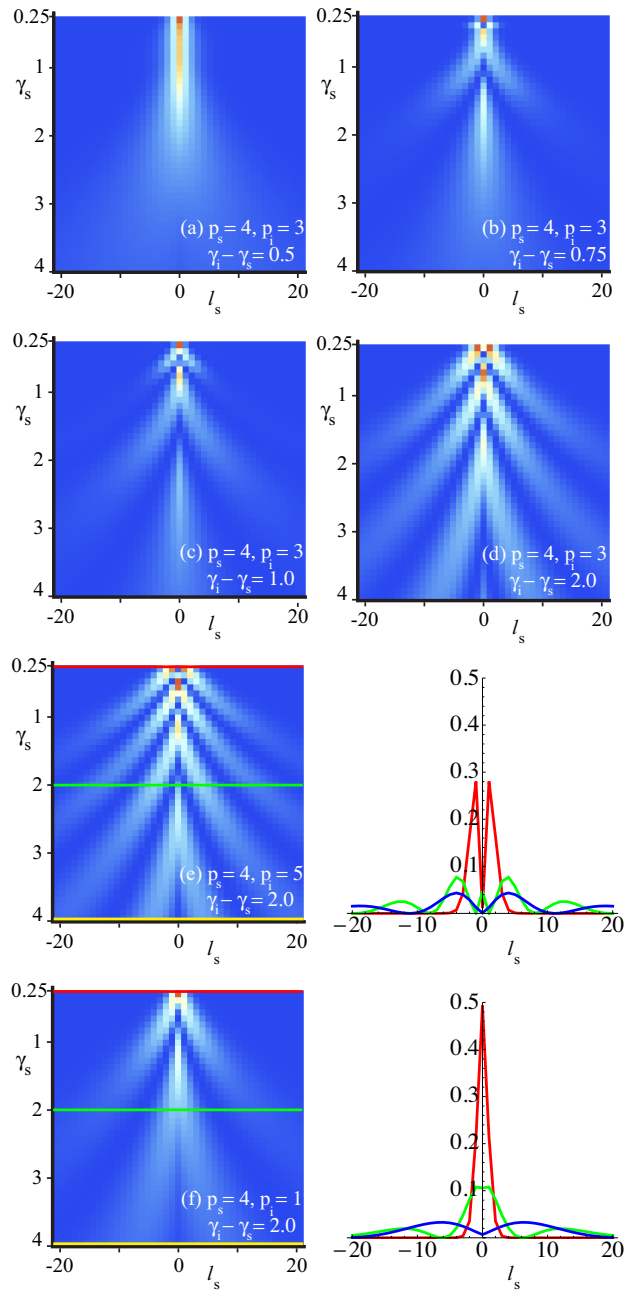}
\caption{(Color online) Joint effect of having a signal and an idler mode with different radial indices and different widths.}
\label{fig:BAND4}
\end{figure}

The physical explanation for the shapes of the SBs is similar to the one supplied in the case of equal radial indices and different signal-idler sizes. The effect is dependent on both the relative size of signal and idler and on the number of rings in the modes. The fact that the number of branches depends on the smaller of $p_i$ and $p_s$ clarifies what is stated above, namely that it all depends on how the rings in the modes overlap: if the joint detection mode consists of two modes with $p_s+1$ and $p_i+1$ rings, the rings will overlap more or less effectively depending on both the relative size and the intensity of the modes. Therefore, it's the mode with less rings that counts.
When $\ell$ is scanned, the two modes, having a different width, will shift the rings away from the center at different rates, giving alternating mostly positive or mostly negative groups of rings. Since the total number of rings depends on the lowest of the radial indices, the mode with less rings will determine the maximum number of branches in the SB, as is shown in Fig.\ref{fig:BAND4}.

\section{Conclusion}
We have calculated the form of the SPDC state produced with a Gaussian pump beam and investigated the 
resulting correlations in the OAM and radial momentum of the down-converted Laguerre-Gaussian beams.
Our results show excellent agreement with previous works on SPDC \cite{Torres03}.

We have shown that, with the usual paraxial and collinear assumptions, the phase matching term can be neglected allowing an analytic expression for the amplitudes to be derived. In addition, we have explored the effect of the radial contributions and of the beams sizes, showing the main families of effects on the radial correlations and on the SB.

In future we will investigate further the phase matching term so that we can consider non-collinear systems. We also will look at the effect of a spatial light modulator on the radial components, since it is the device that is used to convert the LG modes for detection by a single mode fibre.

\section*{Acknowledgements}
We thank Miles Padgett and members of the Optics group at the 
University of Glasgow for useful discussions. 
This work was supported by the UK EPSRC.
We acknowledge the financial support of the Future and Emerging
Technologies (FET) program within the Seventh Framework Programme
for Research of the European Commission, under the FET Open grant
agreement HIDEAS number FP7-ICT-221906.
This research was supported by the DARPA InPho
program through the US Army Research Office award
W911NF-10-1-0395. 
SMB thanks the Royal Society and the Wolfson Foundation for their support.


\end{document}